\documentclass[12pt]{article}
\usepackage{amssymb,amsmath,epsfig}
\usepackage{graphicx}
\usepackage{subfig}
\usepackage{cite}
 \allowdisplaybreaks
\begin{document}
\title{\bf Influence of $f(R)$ Models on the Existence of Anisotropic Self-Gravitating Systems}

\author{Z. Yousaf$^1$ \thanks{zeeshan.math@pu.edu.pk}, M. Sharif$^1$ \thanks{msharif.math@pu.edu.pk}
M. Ilyas$^2$ \thanks{ilyas\_mia@yahoo.com} and M. Z. Bhatti$^1$ \thanks{mzaeem.math@pu.edu.pk}\\
$^1$ Department of Mathematics, University of the Punjab,\\
Quaid-i-Azam Campus, Lahore-54590, Pakistan\\
$^2$ Centre for High Energy Physics, University of the Punjab,\\
Quaid-i-Azam Campus, Lahore-54590, Pakistan}

\date{}

\maketitle

\begin{abstract}
This paper aims to explore some realistic configurations of
anisotropic spherical structures in the background of metric $f(R)$
gravity, where $R$ is the Ricci scalar. The solutions obtained by
Krori and Barua are used to examine the nature of particular compact
stars with three different modified gravity models. The behavior of
material variables is analyzed through plots and the physical
viability of compact stars is investigated through energy conditions. We also
discuss the behavior of different forces, equation of state
parameter, measure of anisotropy and Tolman-Oppenheimer-Volkoff
equation in the modeling of stellar structures. The comparison
from our graphical representations may provide
evidences for the realistic and viable $f(R)$ gravity models at both
theoretical and astrophysical scale.
\end{abstract}

\section{Introduction}

General relativity (GR) laid down the foundations of modern
cosmology. The observational ingredients of $\Lambda$-cold dark
matter model is found to be compatible with all cosmological
outcomes but suffers some discrepancies like cosmic coincidence and
fine-tuning \cite{zs1}. The accelerated expansion of the universe is
strongly manifested after the discovery of unexpected reduction in
the detected energy fluxes coming from cosmic microwave background
radiations, large scale structures, redshift and Supernovae Type Ia
surveys \cite{zs2}. These observations have referred dark energy
(DE) (an enigmatic force), a reason behind this interesting and
puzzling phenomenon. Various techniques have been proposed in order
to modify Einstein gravity in this directions. Qadir et al.
\cite{zs3} discussed various aspects of modified relativistic
dynamics and proposed that GR may need to modify to resolve various
cosmological issues, like quantum gravity and dark matter problem.

The modified theories of gravity are the generalized models came
into being by modifying only the gravitational portion of the GR
action (for further reviews on DE and modified gravity, see, for
instance,~\cite{R2,R3,R4,R5,R6,R7,R8,R9,R10,k3,k4,k5,k6}). The first
theoretical and observationally viable possibility of our
accelerating cosmos from $f(R)$ gravity was proposed by Nojiri and
Odintsov \cite{no1}. There has been interesting discussion of dark cosmic contents on the structure formation
and the dynamics of various celestial bodies in Einstein-$\Lambda$
\cite{kkk1}, $f(R)$ \cite{z1fr}, $f(R,T)$ \cite{z2frt} ($T$ is the
trace of energy momentum tensor) and $f(R,T,R_{\mu\nu}T^{\mu\nu})$
gravity \cite{z3frtrmn}. Recently, Nojiri \emph{et al.} \cite{bin4}
have studied variety of cosmic issues, like early-time, late-time
cosmic acceleration, bouncing cosmology. They emphasized that some
extended gravity theories such as $f(R),~f(G)$ (where $G$ is the
Gauss-Bonnet term) and $f(\mathcal{T})$ (where $\mathcal{T}$ is the
torsion scalar) can be modeled to unveil various interesting cosmic
scenarios.

The search for the effects of anisotropicity in the matter
configurations of compact objects is the basic key leading to
various captivating phenomena, like transitions of phase of
different types \cite{zs4}, condensation of pions \cite{zs5},
existence of a solid as well as Minkowskian core \cite{zs6, zs6a}
etc. One can write all possible exact solutions of static isotropic
relativistic collapsing cylinder in terms of scalar expressions in
GR \cite{zs11} as well as in $f(R)$ gravity \cite{zs12}. Sussman and
Jaime \cite{zs15} analyzed a class of irregular spherical solutions
in the presence of a specific traceless anisotropic pressure tensor
for the choice of $f(R)\propto\sqrt{R}$ model. Shabani and Ziaie \cite{zs17} used
dynamical and numerical techniques to analyze the effects of a
particular $f(R,T)$ gravity model on the stability of emergent
Einstein universe. Garattini and Mandanici \cite{zs18} examined some
stable configurations of various anisotropic relativistic compact
objects and concluded that extra curvature gravitational terms
coming from rainbow's gravity likely to support various patterns of
compact stars. Sahoo and his collaborators \cite{zs19} explored
various cosmological aspects in the context anisotropic relativistic
backgrounds.

Gravitational collapse (GC) is an interesting process, due
to which the stellar bodies could gravitate continuously to move
towards their central points. In this regard, the singularity
theorem \cite{zs20} states that during this implosion process of
massive relativistic structures, the spacetime singularities may
appear in the realm of Einstein's gravity. The investigation of the
final stellar phase has been a source of great interest for many
relativistic astrophysicists and gravitational theorists. In this
respect, various authors \cite{zs23} explored the problem of GC by
taking some realistic configurations of matter and geometry. In the
context of $f(R)$ gravity, various results have been found in
literature about collapsing stellar interiors and black holes
\cite{zs24}.

Capozziello et al. \cite{zs25} studied GC of non-interacting
particles by evaluating dispersion expressions through perturbation
approach and found some unstable regime of the collapsing object
under certain limits. Cembranos et al. \cite{zs26} examined GC of
non-static inhomogeneous gravitational sources and studied the
large-scale structure formation at early-time cosmic in different
$f(R)$ gravity theories. Modified gravity theories likely to host
massive celestial objects with smaller radii as compared to GR
\cite{zs31}. Guo and Joshi \cite{zs40} discussed scalar GC of
spherically symmetric spacetime and inferred that relativistic
sphere could give rise to black hole configurations, if the source
field is strong.

The concept of energy conditions (ECs) could be considered as
viable approach for the better understanding of the well-known
singularity theorem. Santos et al. \cite{zs33} developed viability bounds coming from ECs
on generic $f(R)$ formalism. Their approach could be considered to
constrain various possible $f(R)$ gravity models with proper
physical backgrounds.  Wang et al. \cite{zs34} evaluated some
generic expressions for ECs in the $f(R)$ gravity and
employed them on a class of cosmological model to obtain some
viability constraints.

Shiravand et al. \cite{zs35} evaluated ECs for $f(R)$ gravity and
obtained some stability constraints against Dolgov-Kawasaki
instability. They found special ranges of some $f(R)$ model
parameters under which the theory would satisfy WECs. The
investigation of ECs in modified theories has been carried out under
a variety of cosmological issues like, $f(R)$ gravity \cite{zs36},
$f(R,L_m)$ gravity \cite{zs37}, $f(R,T)$ gravity \cite{zs38},
$f(R,G)$ gravity \cite{zs39}, $f(G)$ gravity \cite{kk37}. The
stability of compact objects along with their ECs have been analyzed
in detail by various researchers \cite{zs42}.

The main goal of this paper is to investigate the role of $f(R)$
models as well as anisotropic pressure in the modeling of
realistic compact stellar structures. We study various structural
properties, like distributions of density and pressure
anisotropicity, Tolman-Oppenheimer-Volkoff (TOV) equation, energy
conditions, stability as well as equation of state parameter, for
three different observational data of stellar structures. The paper
is outlined as follows. In the next section, we
briefly review $f(R)$ gravity for anisotropic
configurations of the static spherical geometry. Section \textbf{3}
is devoted to demonstrate some $f(R)$ gravity models along with
their physical viability. In section \textbf{4}, we check
physical viability of three well-known stellar structure. At the
end, we conclude our main findings.

\section{Anisotropic Relativistic Spheres in $f(R)$ Gravity}

The standard Einstein-Hilbert action in $f(R)$ gravity can be
modified as follows
\begin{equation}\label{1}
S_{f(R)}=\frac{1}{2\kappa}\int d^4x\sqrt{-g}f(R)+S_M,
\end{equation}
where $g,~\kappa,~S_M$ stand for the determinant of the tensor,
the coupling constant and matter field action.
The basic motivation of this theory is to introduce generic
algebraic expression of the Ricci scalar rather than cosmological
constant in the GR action. By giving variation in the above equation
with respect to $g_{\mu\nu}$, the field equations for $f(R)$ gravity
can be found as
\begin{equation}\label{2}
R_{\alpha\beta}f_R-\frac{1}{2}f(R)g_{\alpha\beta}+\left(g_{\alpha\beta}{\Box}
-\nabla_{\alpha}\nabla_{\beta}\right)f_R={\kappa}T_{\alpha\beta},
\end{equation}
where $T_{\alpha\beta}$ is the standard
energy-momentum tensor, while $\nabla_{\beta}$ is an operator of covariant derivative,
$\Box\equiv \nabla^{\beta}\nabla_{\beta}$ and $f_R\equiv df/dR$. The
quantity $f_R$ comprises of second corresponding derivatives of the
metric variables, which is often termed as scalaron which propagates
new scalar freedom degrees. The trace of Eq.(\ref{2}) specifies
scalaron equation of motion as under
\begin{align}\label{3}
3\Box{f_R}+{R}f_R-2f(R)={\kappa}T,
\end{align}
where $T\equiv T^\beta_{\beta}$. It has been noticed that above
equation is a second order differential equation in $f_R$, unlike GR
in which the trace of Einstein field equation boils down to
$R=-\kappa T$. This points $f_R$ as a source of producing scalar
degrees of freedom in $f(R)$ theory. The condition $T=0$ does not
necessarily implies the vanishing (or constant value) of the Ricci
scalar in the dynamics. This presents Eq.(\ref{3}) as a useful
mathematical tool to discuss many hidden and interesting cosmic
arena, like Newtonian limit, stability etc. The constraint, constant
Ricci scalar as well as $T_{\alpha\beta}=0$, boil down Eq.(\ref{3}) as
\begin{align}
Rf_R-2f(R)=0,
\end{align}
which is the Ricci algebraic equation after choosing any viable
formulations of $f(R)$ model. If someone finds a constant roots of
the above equation, i.e., $R=\Lambda$ (say), then Eq.(\ref{3}) yields
\begin{align}
R_{\alpha\beta}=\frac{g_{\alpha\beta}\Lambda}{4},
\end{align}
thereby indicating (anti) de Sitter as the maximally symmetric
solution. Equation (\ref{2}) can be remanipulated as
\begin{equation}\label{4}
G_{\alpha\beta}=\frac{\kappa}{f_R}(\overset{(D)}
{T_{\alpha\beta}}+T_{\alpha\beta})\equiv {T_{\alpha\beta}^{\textrm{eff}}},
\end{equation}
where $G_{\alpha\beta}$ is an Einstein tensor and $\overset{(D)}
{T_{\alpha\beta}}$ is termed as effective form of the
energy-momentum tensor. Its expression is given by
\begin{equation*}
\overset{(D)}{T_{\alpha\beta}}=\frac{1}{\kappa}\left\{
\nabla_{\alpha}\nabla_{\beta}f_R-\Box
f_Rg_{\alpha\beta}+(f-Rf_R)\frac{g_{\alpha\beta}}{2}\right\}.
\end{equation*}

We consider a general form of a static spherically symmetric line
element as
\begin{equation}\label{z7}
d{s^2} = e^{a}d{t^2} -e^{b}d{r^2}-{r^2}( {d{\theta ^2} + {{r^2}{\sin }^2}\theta d{\phi ^2}}),
\end{equation}
where $a$ and $b$ are radial dependent metric coefficients. We
assume that our spherical self-gravitating system is filled with
locally anisotropic relativistic fluid distributions. The
energy-momentum tensor of this matter is given by
\begin{equation}\label{z8}
{T_{\mu \nu }} = (\rho  + p_t){U_\mu }{U_\nu } - {p_t}{g_{\mu \nu }} + ({p_r} - {p_t}){V_\mu }{V_\nu },
\end{equation}
where $p_r,~p_t$ are the radial and tangential components of
pressure and $\rho$ is the fluid energy density. The four-vectors
$U_\mu$ and $V_\mu$, under non-tilted coordinate frame, obey
relations $U_\mu U^\mu=1$ and $V_\mu V^\mu=-1$. The $f(R)$ field
equations (\ref{4}) for the metric (\ref{z7}) and fluid (\ref{z8})
can be given as
\begin{align}\nonumber
\rho &= \frac{{{e^{ - b}}}}{{2{r^2}}}({r^2}b'{f_R}^\prime +
2{f_R}rb' + f( - {r^2}{e^{b}}) + {f_R}{r^2}{e^{b}}R + 2{f_R}{e^{b}}
- 2{r^2}{f_R}^{\prime \prime }\\\label{z1} & - 4r{f_R}^\prime -
2{f_R}),\\\label{z2} {p_r}&=  - \frac{{{{\rm{e}}^{ -
b}}}}{{2{r^2}}}( - 2{f_R} + 2{{\rm{e}}^b}{f_R} - {{\rm{e}}^b}f{r^2}
+ {{\rm{e}}^b}{f_R}{r^2}R- 2{f_R}ra' - 4r{f_R}^\prime-
{r^2}a'{f_R}^\prime ),\\\nonumber {p_t}&= \frac{{{e^{ -
b}}}}{{4r}}(2{f_R}ra'' - {f_R}ra'b' + 2ra'{f_R}^\prime +
{f_R}r{{a'}^2} + 2{f_R}a' - 2rb'{f_R}^\prime - 2{f_R}b'\\\label{z3}
& + 2rf{e^b} - 2r{f_R}{e^b}R + 4r{f_R}^{\prime \prime } +
4{f_R}^\prime ).
\end{align}
The corresponding Ricci scalar is given by
\begin{equation}
R = \frac{{{{\rm{e}}^{ - b}}}}{{2{r^2}}}\left[4 - 4{{\rm{e}}^b}
+ {r^2}{a'^2} - 4rb' + ra'\left( {4 - rb'} \right) + 2{r^2}{a^{\prime \prime }}\right].
\end{equation}
In order to achieve some realistic study for the modeling of
anisotropic compact stellar structure, we use a specific
combination of metric variables, i.e., $a(r) =B r^2 +C$ and $b(r)=A
r^2$ suggested by Krori and Barua \cite{zs43}. Here, $A,~B$ and $C$
are constants and can be found by imposing some viable physical
grounds. Making use of these expressions, the $f(R)$ field equations
can be recasted as
\begin{align}\nonumber
\rho &= \frac{1}{{2{r^2}}}{{\rm{e}}^{ - {r^2}A}}( - 2{f_R} +
2{{\rm{e}}^{{r^2}A}}{f_R} - {{\rm{e}}^{{r^2}A}}f{r^2} + 4{f_R}{r^2}A
+{{\rm{e}}^{{r^2}A}}{f_R}{r^2}R +2{f_R}{r^3}A' -
4r{f_R}^\prime\\\label{ro} & + 2{r^3}A{f_R}^\prime  +
{r^4}A'{f_R}^\prime  - 2{r^2}{f_R}^{\prime \prime }),\\\nonumber
{p_r}&=  - \frac{1}{{2{r^2}}}{{\rm{e}}^{ - {r^2}A}}( - 2{f_R} +
2{{\rm{e}}^{{r^2}A}}{f_R} - {{\rm{e}}^{{r^2}A}}f{r^2} - 4{f_R}{r^2}B
+ {{\rm{e}}^{{r^2}A}}{f_R}{r^2}R - 2{f_R}{r^3}B'\\\label{pr} & -
4r{f_R}^\prime  - 2{r^3}B{f_R}^\prime  - {r^4}B'{f_R}^\prime
),\\\nonumber {p_t}& = \frac{1}{{4r}}{{\rm{e}}^{ -
{r^2}A}}(2{{\rm{e}}^{{r^2}A}}fr - 4{f_R}rA + 8{f_R}rB -
4{f_R}{r^3}AB + 4{f_R}{r^3}{B^2} - 2{{\rm{e}}^{{r^2}A}}{f_R}rR -
2{f_R}{r^2} \\\nonumber &\times A'- 2{f_R}{r^4}BA' + 10{f_R}{r^2}B'
- 2{f_R}{r^4}AB'+ 4{f_R}{r^4}BB' - {f_R}{r^5}A'B' +
{f_R}{r^5}{{B'}^2} + 4{f_R}^\prime   \\\label{pt} &- 4{r^2}
A{f_R}^\prime + 4{r^2}B{f_R}^\prime  - 2{r^3}A'{f_R}^\prime +
2{r^3}B'{f_R}^\prime  + 2{f_R}{r^3}{B^{\prime \prime }} +
4r{f_R}^{\prime \prime }).
\end{align}

Now, we consider a three-dimensional hypersurface, $\Sigma$ that has
differentiated our system into interior and exterior regions. The
spacetime for the description of exterior geometry is given by the
following vacuum solution
\begin{equation}
d{s^2} = \left( {1 - \frac{{2M}}{r}} \right)d{t^2}
- {\left( {1 - \frac{{2M}}{r}} \right)^{ - 1}}d{r^2}
- {r^2}\left( {d{\theta ^2} + {\mathop{\rm \sin}\nolimits} {\theta ^2}d{\varphi ^2}} \right),
\end{equation}
where $M$ is the gravitating mass of the black hole. The continuity
of the structural variables, i.e., $g_{ii},~i=1,2$ and
derivatives $\frac{\partial g_{tt}}{\partial r}$ over the
hypersurface, i.e., $r=R$, provide some equations. On solving these
equations simultaneously, one can obtain
\begin{align}
A& = \frac{{ - 1}}{{{R^2}}}\ln \left( {1 - \frac{{2M}}{r}} \right),\quad
B = \frac{M}{{{R^3}}}{\left( {1 - \frac{{2M}}{r}} \right)^{ - 1}},\\
C &= \ln \left( {1 - \frac{{2M}}{r}} \right) - \frac{M}{R}{\left( {1 - \frac{{2M}}{r}} \right)^{ - 1}}.
\end{align}
After selecting some particular values of $M$ and $R$, the
corresponding values of metric coefficients $A$ and $B$ can be
found. Some possibilities of such types are mentioned in Table
\ref{table:1}.
\begin{table}[h!]
\caption{The approximate values of the masses $M$, radii $R$,
compactness $\mu$, and the constants $A$ and $B$ for the compact
stars, Her X-1, SAXJ 1808.4-3658, and 4U 1820-30.} \label{table:1}
\centering
\begin{tabular}{|c| c| c| c| c| c|}
\hline Compact Stars  &$M$ & $R(km)$ & $\mu=\frac{M}{R}$
&$A(km^{-2})$ &$B(km^{-2})$ \\ [0.5ex] \hline\hline\ Her X-1  &
$0.88M_{\odot}$ & $7.7$ & $0.168$ & $0.006906276428 $
&$0.004267364618 $\\ [1ex] \hline\ SAXJ1808.4-3658& $1.435M_{\odot}$
& $7.07$ & $0.299$ & $0.01823156974$ &$0.01488011569$\\ [1ex] \hline
4U1820-30& $2.25M_{\odot}$ & $10$ & $0.332$ & $0.01090644119 $
&$0.009880952381$\\ [1ex] \hline
\end{tabular}
\end{table}

\section{Physical Aspects of $f(R)$ Gravity Models}

In this section, we consider some well-known viable $f(R)$ models of
gravity for the description of some physical environment of compact
stellar interiors. We shall check evolution of energy density,
pressure, equation of state parameter, TOV equation and energy
conditions for some particular stars with three above mentioned
$f(R)$ models. We use three configurations of stellar bodies, i.e.,
Her X-1, SAX J 1808.4-3658, and 4U 1820-30 of masses
$0.88M_{\odot},~1.435M_{\odot}$ and $2.25M_{\odot}$, respectively.
We use three
$f(R)$ models in Eqs.(\ref{ro})-(\ref{pt}) to obtain
$\rho,~p_r$ and $p_t$. These equations would assist us to
investigate various stability features of compact stellar structures
(shown in Table \ref{table:1}). We shall also check the
corresponding behavior of stellar interiors by drawing plots. In
the diagrams, the stellar structures Her X-1, SAX J 1808.4-3658, and
4U 1820-30 are labeled with CS1, CS2 and CS3 abbreviations,
respectively.

The strange stars are widely known as those quark structures that
are filled with strange quark matter contents. There has been an
interesting theoretical evidences that indicate that quark stars
could came into their existence from the remnants of neutron stars
and forceful supernovas \cite{50s}. Surveys suggested the possible
existence of such structures in the early epochs of cosmic history
followed by the Big Bang \cite{51s}. On the other hand, the evolution of
stellar structure could end up with white dwarfs, neutron stars or
black holes, depending upon their initial mass configurations. Such
structures are collectively dubbed with the terminology, compact
stars.

A maximum permitted mass radius ratio for the case of static
background of spherical relativistic structures coupled with ideal
matter distributions should be $2M/R<\frac{8}{9}$. This result has
gained certain attraction among relativistic astrophysicists in
order to design the existence of compact structures and is widely
known as Buchdahl-Bondi bound \cite{52s, 53s}. In this paper, we consider
the radial and transverse sound speed in order to perform stability
analysis. However, in order to investigate the equilibrium conditions, we shall study the impact of hydrostatic, gravitational and 
anisotropic forces, in the possible modeling of compact stars.

\subsection{Model 1}

Firstly, we assume model in power-law form of the Ricci scalar given
by \cite{zs44}
\begin{equation}\label{model1}
f(R) = R + \alpha {R^2},
\end{equation}
where $\alpha$ is a constant number. Starobinsky presented this
model for highlighting the exponential growth of early-time cosmic
expansion. This Ricci scalar formulation, in most manuscripts, is
introduced for the possible candidate of DE. Einstein's gravity
introduced can be retrieved, under the limit $f(R)\rightarrow R$.

\subsection{Model 2}

Next, we consider Ricci scalar exponential gravity given by
\cite{zs45}
\begin{equation}\label{model2}
f = R + \beta R\left( {{e^{\left( { - R/\tilde{R} } \right)}} - 1} \right),
\end{equation}
where $\beta$ and $\tilde{R}$ are constants. The models of such
configurations has been studied in the field of cosmology by
\cite{zs46}. The consideration of this model could provide an active
platform for the investigation of late-time accelerating universe
complying with matter-dominated eras.

\subsection{Model 3}

It would be interesting to consider $f(R)$ corrections of the form
\begin{equation}\label{model3}
f = R + \alpha {R^2}\left( {1 + \gamma R} \right),
\end{equation}
in which $\alpha$ and $\gamma$ are the arbitrary constant. The
constraint $\gamma R\sim \mathcal{O}(1)$ specifies this model
relatively more interesting as one can compare their analysis of
cubic Ricci scalar corrections with that of quadratic Ricci term.

\subsection{Energy Density and Pressure Evolutions}

\begin{figure} \centering
\epsfig{file=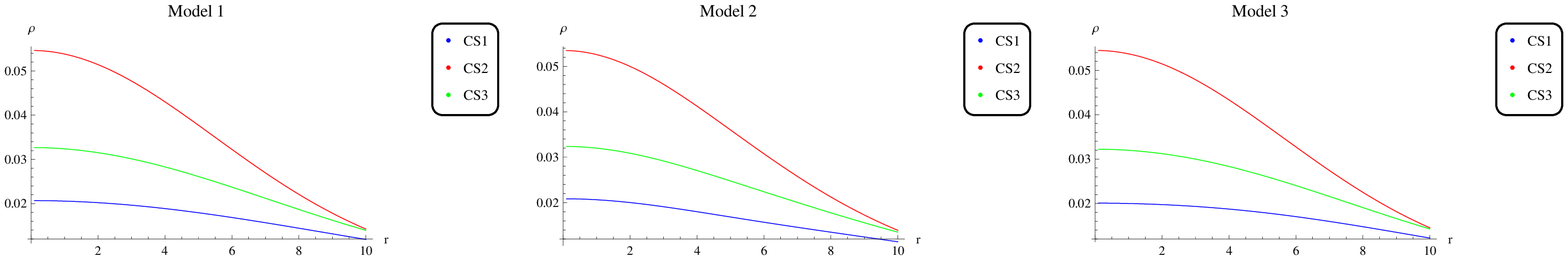,width=1\linewidth} \caption{Density evolution of
the strange star candidates with three different $f(R)$
models.}\label{roc} \epsfig{file=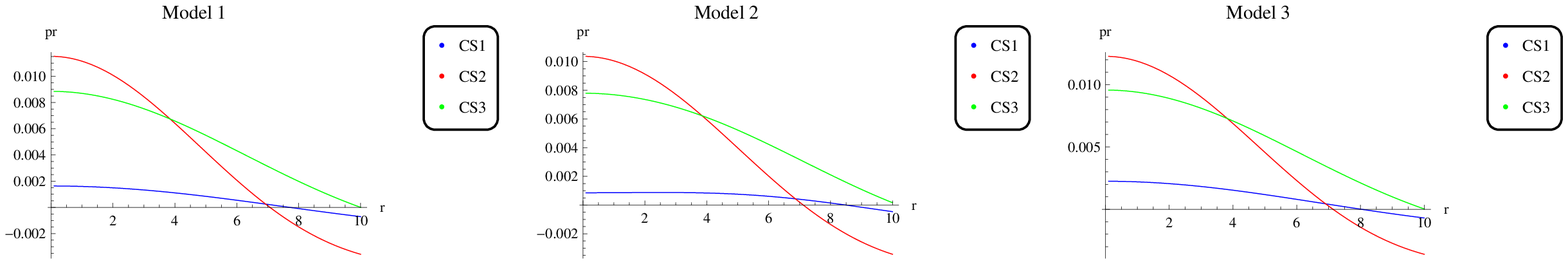,width=1\linewidth}
\caption{Radial pressure evolution of the strange stars.}\label{prc}
\epsfig{file=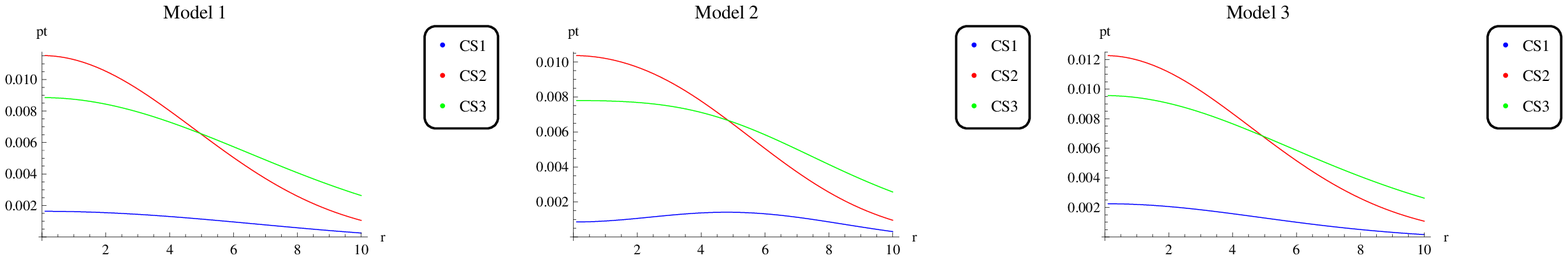,width=1\linewidth} \caption{Transverse pressure
evolution of the strange stars.}\label{ptc}
\end{figure}

\begin{figure} \centering
\epsfig{file=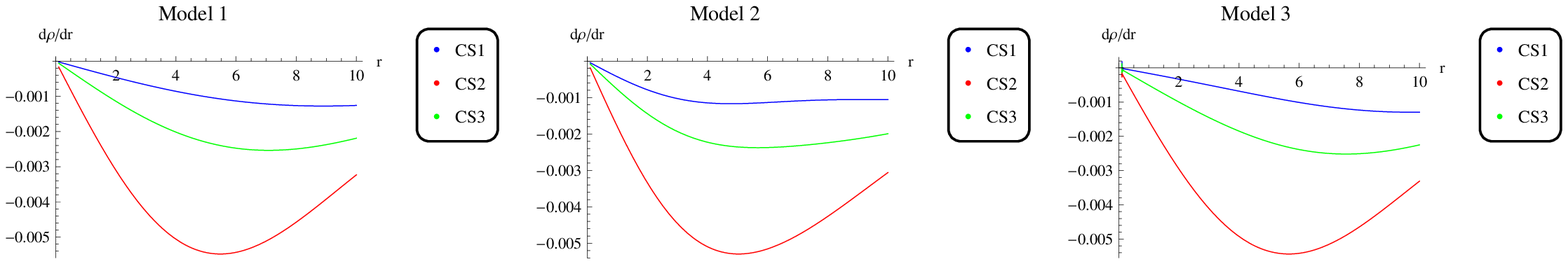,width=1\linewidth} \caption{Behavior of
$d\rho/dr$ with respect to $r$ with three different $f(R)$
models.}\label{dro} \epsfig{file=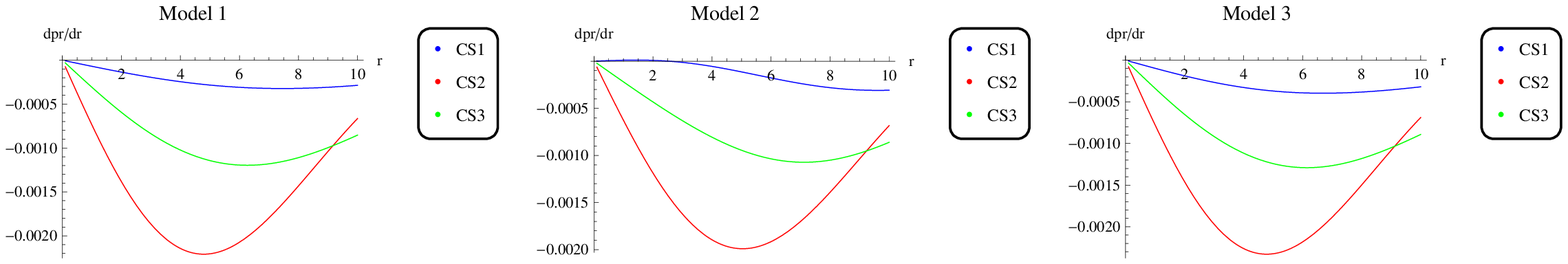,width=1\linewidth}
\caption{Behavior of $dp_r/dr$ with respect to $r$.}\label{dpr}
\epsfig{file=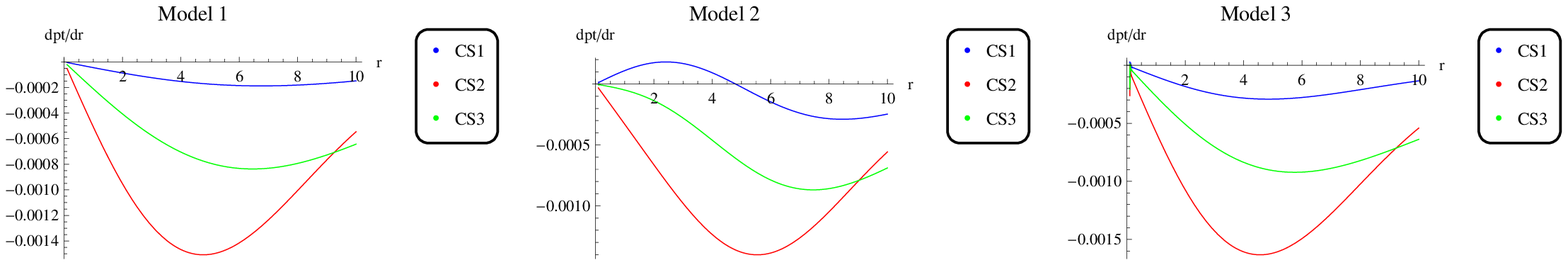,width=1\linewidth} \caption{Behavior of
$dp_t/dr$ with increasing $r$.}\label{dpt}
\end{figure}

Here, we analyze the domination of star matter as well as
anisotropic pressure at the center with $f(R)$ models. The
corresponding changes in the profiles of energy density, radial and
transverse pressures are shown in Figures (\ref{dro})-(\ref{dpt}),
respectively. We see that $\frac{d\rho}{{dr}}<0$,
$\frac{dp_r}{{dr}}<0$ and $\frac{dp_t}{{dr}}<0$ for all three models
and strange stars. For $r=0$, we obtain
$$\frac{d\rho}{{dr}}=0,\quad \frac{dp_r}{{dr}}=0,$$
which is expected because these are monotonically decreasing
functions. One can observe maximum impact of density (star core
density $\rho(0)=\rho_c$) for small $r$. The plot of the density for
the strange star candidate Her X-1, SAX J 1808.4-3658, and 4U
1820-30 are drawn. Figure (\ref{roc}) shows that as $r\rightarrow0$,
the density $\rho$ profile keeps on increasing its value, thereby
indicating $\rho$ as the monotonically decreasing function of $r$.
This suggests that $\rho$ would decrease its effects on
increasing $r$ which indicates high compactness at
the stellar core. This proposes that our chosen $f(R)$ models may
provide viable results at the outer region of the core. The other
two plots, shown in Figures (\ref{prc}) and (\ref{ptc}), indicate
the variations of the anisotropic radial and transverse pressure,
$p_r$ and $p_t$.

\subsection{Energy Conditions}

\begin{figure}
\centering \epsfig{file=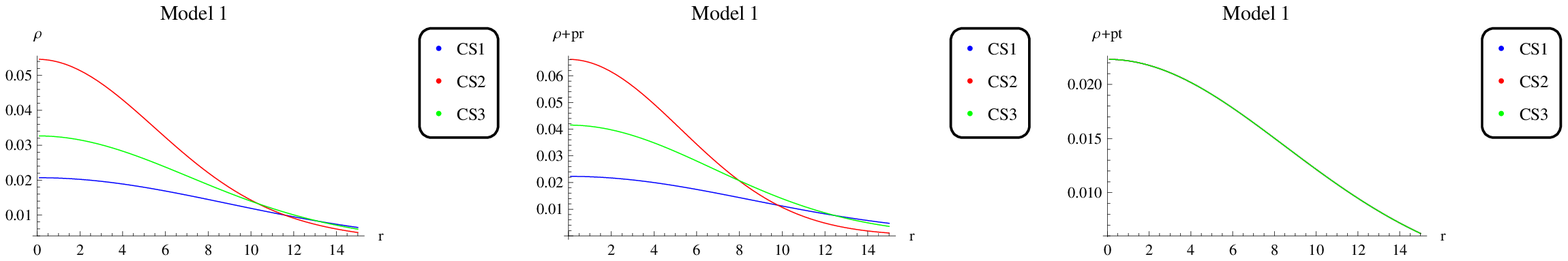,width=1\linewidth}
\epsfig{file=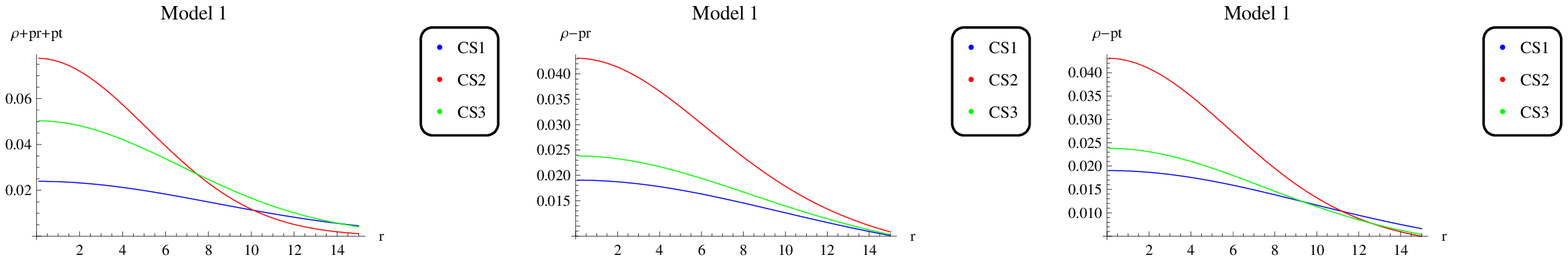,width=1\linewidth} \caption{Viability of
energy conditions for three different strange stars with dark source
terms coming from $f(R) = R + \alpha {R^2}$.}\label{energy1}
\end{figure}

\begin{figure}
\centering \epsfig{file=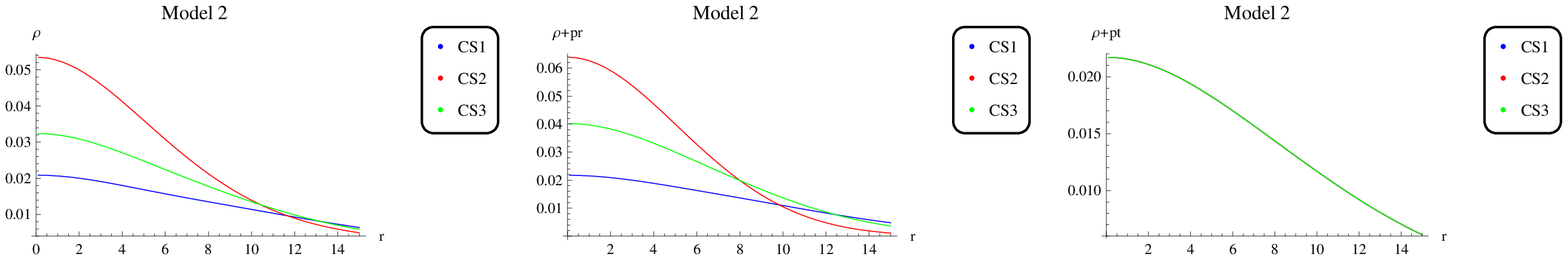,width=1\linewidth}
\epsfig{file=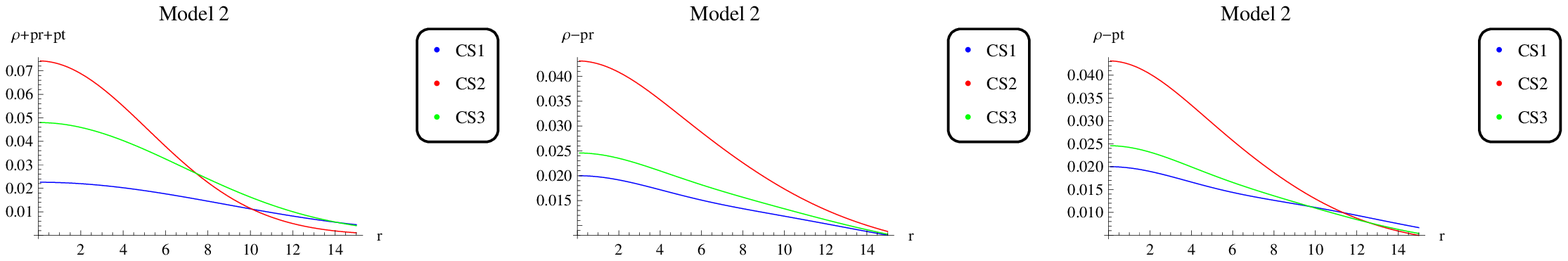,width=1\linewidth} \caption{Viability of
energy conditions for three different strange stars with $f(R) = R +
\beta R\left( {{e^{\left( { - R/\tilde{R} } \right)}} - 1} \right)$
model.}\label{energy2}
\end{figure}

\begin{figure}
\centering \epsfig{file=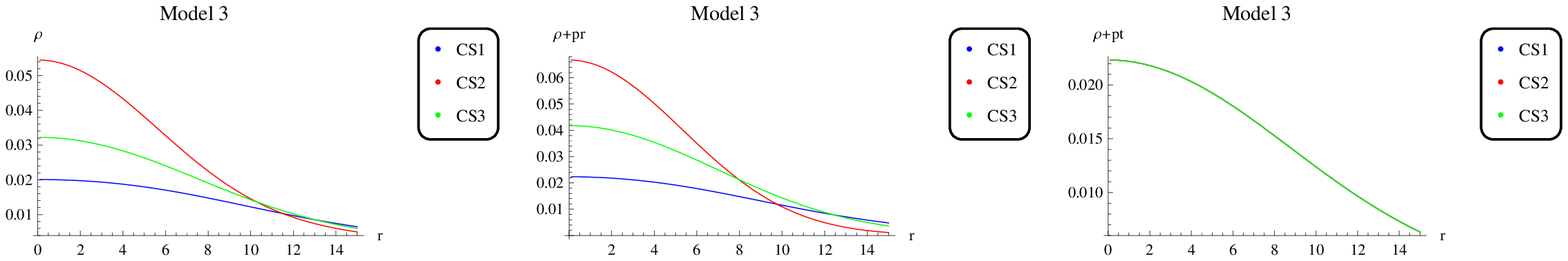,width=1\linewidth}
\epsfig{file=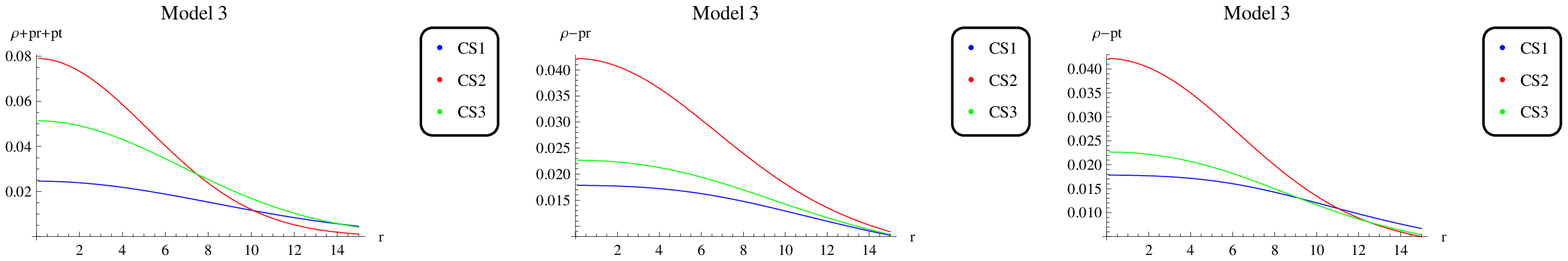,width=1\linewidth} \caption{Viability of
energy conditions for three different strange stars in the presence
of $f(R) = R + \alpha {R^2}\left( {1 + \gamma R} \right)$
corrections.}\label{energy3}
\end{figure}

For the mathematical expressions of stress-energy tensors to
represent physically acceptable matter fields, it should obey some
particular constraints, widely known as ECs. These ECs have very
important property that they are coordinate-invariant (independent
of symmetry). Recently, Yousaf \cite{zs47} explored various
analytical models of the stellar filaments with dark source terms
coming from cosmological constant and checked the validity regimes
of ECs in order to make them physically viable. Bamba et al.
\cite{zs48} presented general formalism for checking the viability
of ECs in modified gravity. In $f(R)$ gravity (having effective
density and anisotropic pressure), the NEC and WEC are formulated as
\begin{align}\nonumber
&\textrm{NEC}\Leftrightarrow {\rho^{\textrm{eff}}} +
{p_i^{\textrm{eff}}} \ge 0,\\\nonumber &\textrm{WEC}\Leftrightarrow
{\rho^{\textrm{eff}}} \ge 0 \text{ and } {\rho^{\textrm{eff}}} +
{p_i^{\textrm{eff}}} \ge 0,
\end{align}
while the SEC and the dominant energy condition (DEC) yield
\begin{align}\nonumber
&\textrm{SEC}\Leftrightarrow {\rho^{\textrm{eff}}+
3{p_r^{\textrm{eff}}}}+2{p_t^{\textrm{eff}}} \ge 0 \text{ and } {\rho^{\textrm{eff}}} +
{p_i^{\textrm{eff}}} \ge 0,\\\nonumber &\textrm{DEC} \Leftrightarrow
{\rho^{\textrm{eff}}} \ge 0 \text{ and } {\rho^{\textrm{eff}}} \pm
{p_i^{\textrm{eff}}} \ge 0.
\end{align}
The evolution of all these ECs for three different compact
structures are being well satisfied for all of our chosen $f(R)$
models. These are graphically shown in Figures
(\ref{energy1})-(\ref{energy3}).

\subsection{TOV Equation}

\begin{figure} \centering
\epsfig{file=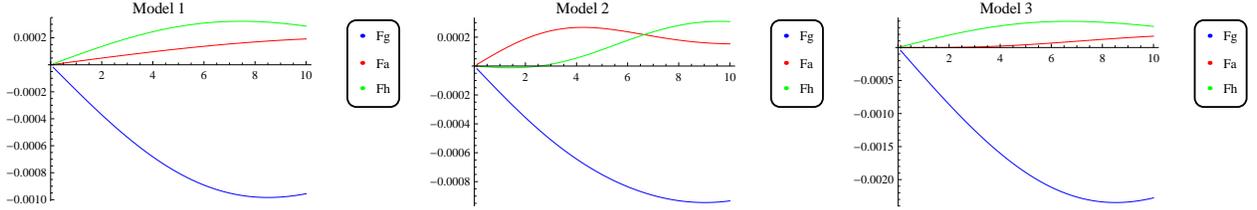,width=1\linewidth} \caption{The plot of
gravitational force ($F_g$), hydrostatic force ($F_h$) and
anisotropic force $(F_a)$ with respect to the radial coordinate
r(km) and three different models.}\label{eqb}
\end{figure}

The TOV equation for the spherical anisotropic stellar interior is
given by
\begin{equation}\label{aa1}
\frac{{d{p_r}}}{{dr}} + \frac{{a'(\rho  + {p_r})}}{2} + \frac{{2({p_r} - {p_t})}}{r} = 0.
\end{equation}
The quantity $a'$ is the radial derivative of the function appearing
in the first metric coefficient of the line element (7). However,
the quantity $a$, in general, is directly correspond to the scalar
associated with the four acceleration $(a_\beta=aV_\beta)$ of the
anisotropic fluid and is defined as
$$a^2=a_\beta a^\beta$$
Equation (\ref{aa1}) can be relabeled in terms of gravitational $(F_g)$,
hydrostatic $(F_h)$ and anisotropic $(F_a)$ forces as
\begin{equation}
F_g + F_h + F_a= 0.
\end{equation}
The values of these forces for our anisotropic spherical matter
distribution have been found as follows
\begin{equation}\nonumber
F_g=-B r (\rho+p_r), \quad F_h=-\frac{{d{p_r}}}{{dr}},\quad F_a= \frac{{2({p_r} - {p_t})}}{r},
\end{equation}
By making use of these definitions, the behavior of these forces for
the onset of hydrostatic equilibrium are shown for three strange
compact stars in Figure (\ref{eqb}). In the graphs, we have
continued our analysis with different $f(R)$ theories. In Figure
(\ref{eqb}), the left plot is showing variations due to model 1,
middle is for model 2 and the right plot is describing corresponding
changes in principal forces due to model 3.

\subsection{Stability Analysis}

\begin{figure} \centering
\epsfig{file=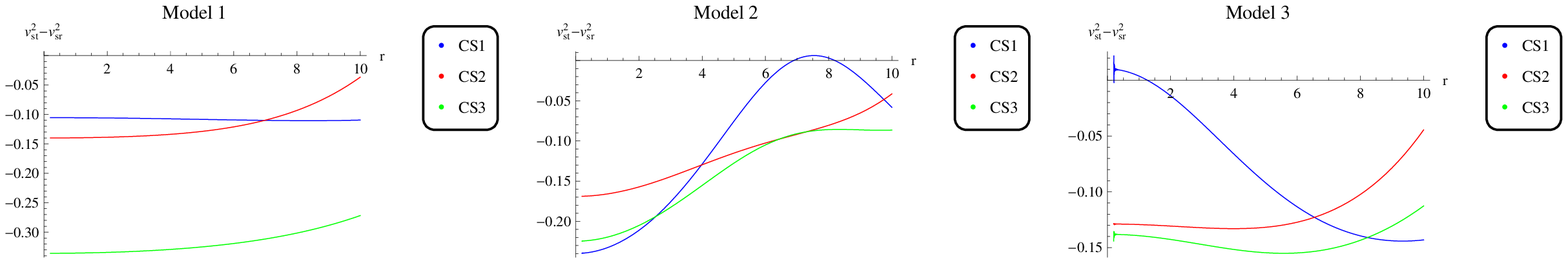,width=1\linewidth}
\caption{Variations of $v_{st}^2 - v_{sr}^2$ with respect radius, $r$ (km)}\label{vstmvsr}
\end{figure}

Here, we check the stability of our stars
by adopting the scheme presented by Herrera \cite{zs49} that was
based on the concept of cracking (or overturning). This approach
states that $v^2_{sr}$ as well as $v^2_{st}$ must belong to the closed
interval [0,1], where $v_{sr}$ indicates radial sound speed, while
and $v_{st}$ denotes transverse sound speed defined as
$$\frac{{d{p_r}}}{{d\rho }} = v_{sr}^2,\quad \frac{{d{p_t}}}{{d\rho }} = v_{st}^2.$$
The system will be dynamically stable, if $v^2_{st}>v^2_{sr}$. It
has been observed that the evolution of the radial and transversal sound speeds for all three
types of strange stars are within the bounds of stability for some
regions. It can be seen from Figure (\ref{vstmvsr}) that all of our
stellar structures (within the background of all $f(R)$ models) obey
the following constraint
$$0< |v_{st}^2-v_{sr}^2|<1.$$
Therefore, we infer that all of our proposed model are stable in
this theory. Such kind of results have been proposed by Sharif and Yousaf \cite{ps1}
by employing different mathematical strategy on compact stellar objects.

\subsection{Equation of State Parameter}

\begin{figure} \centering
\epsfig{file=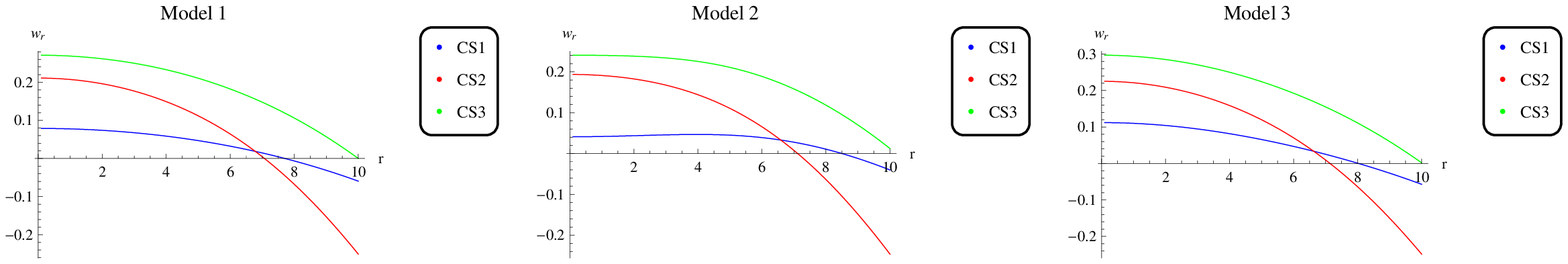,width=1\linewidth} \caption{Variations of the
radial EoS parameter, $\omega_r$ with respect to the radial
coordinate.}\label{wr} 
\epsfig{file=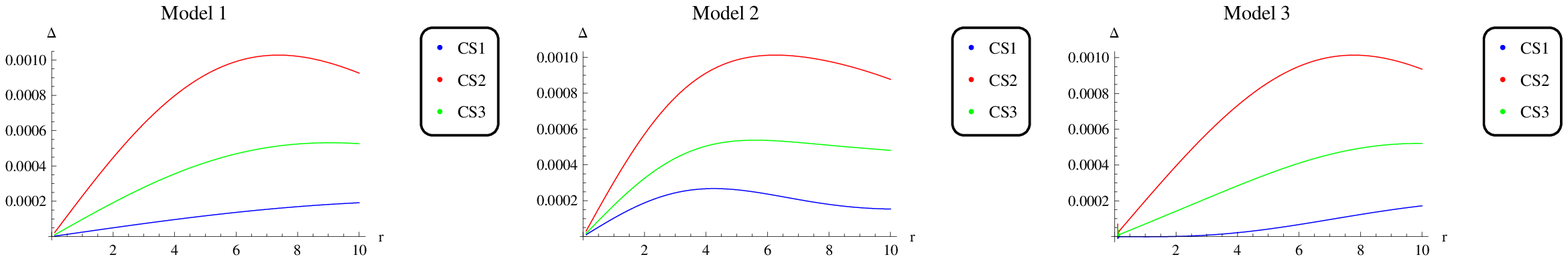,width=1\linewidth} \caption{Variations of
anisotropicity, $\Delta$, with respect to $r$.}\label{ani}
\end{figure}

The parameter corresponding to the equation of state (EoS) is a
dimensionless term that illustrates matter state under some specific
physical grounds. This parameter has its range belonging to open
interval $(0,1)$. In that case, it represents radiation dominated
cosmic era. For anisotropic relativistic interior, the EoS can be
defined as
\begin{align}\nonumber
p_r=\omega_r \rho, \quad p_t=\omega_t \rho.
\end{align}
The behaviors of $\omega_r$ for our compact
structures is shown graphically in Fig.(\ref{wr}). However, the similar behavior of $\omega_t$ 
can be observed for all of our observed compact structures very easily. It has been observed that maximum radius of compact
objects to achieve the limit $0<\omega_r<1$ is $r\sim (\leq7)$,
while the constraint $0<\omega_t<1$ is valid for any large value of
$r$. This means that $\omega_i>1$ near its central point. The
spherically symmetric self-gravitating system would be in a
radiation window at the corresponding hypersurfaces. From here, we
conclude that that our relativistic bodies have a compact interior.

\subsection{The Measurement of Anisotropy}

Here, we measure the extent of anisotropy in the modeling of
relativistic interiors. It is well-known that anisotropicity in
the stellar system can be measured with the help of the following
formula
\begin{equation}
\Delta  = \frac{2}{r}({p_t} - {p_r}),
\end{equation}
The quantity $\Delta$ is directly related to the difference
$p_t-p_r$. The positivity of $\Delta$ indicates the positivity of
$p_t-p_r$. Such a background suggests the outwardly drawn behavior
of anisotropic pressure. However, the resultant pressure will be
directly inward, once $\Delta$ is less than zero. We have drawn the
anisotropic factor for our systems and obtain $\Delta>0$,
thereby giving $p_t > p_r$. All these results are mentioned
through plots as shown in Figure (\ref{ani}).

\section{Summary}

This paper is devoted to discuss some physical aspects of
spherically symmetric compact stars in the background of metric
$f(R)$ theory. We have used the
Krori-Barua solutions for metric functions of spherical star whose
arbitrary constants are explored over the boundary surface by
matching it with suitable exterior. The arbitrary constants of the
Krori-Barua solutions can be written in the form of mass and radius
for any compact star. We have used the observational data of three
particular star models to explore the influence of extra degrees of freedom
on compact stars. For this purpose, three different
physically viable $f(R)$ models are used. By using the values of
these star and gravity models, we have plotted the material
variables like energy density and anisotropic stresses against
radial distance. It is found that as the radius of the star
increases, the density tends to decrease, thereby indicating the
maximum dense configurations of stellar interiors. A similar
behavior is analyzed in the evolutionary phases of the tangential
and radial pressures.

It is seen that $r-$derivatives of these material variables
remains negative with the increasing radius for all the three models
while only the $r-$derivative of tangential pressure for compact
star Her X-1 has some positive value till $r=4.8$ and then becomes
negative. It is also significant to note that the first derivative
of these material variables vanish at $r=0$ for all the compact
stars. It is found that our spherically symmetric anisotropic
systems are obey NECs, WECs, SECs and DECs, hence the
compact stars under the effects of extra degrees of freedom of
fourth order gravity are physically valid. We have seen that the
gravitational forces are overcoming the corresponding repulsive
forces, thus indicating the collapsing nature of compact
relativistic structures. It is well-known that a stellar system
would be stable against fluctuations, if it satisfies the bounds of
[0,1] for radial and tangential sound speeds. We have found that the
compact star Her X-1 under the effects of second gravity model does
not remain in [0,1], but Figure \ref{vstmvsr} indicates that all of our stellar
models are stable (i.e., $v_{st}^2>v_{sr}^2$). Further, we have
observed that the equation of state parameter lies in the interval
(0,1) for all the compact stars. The anisotropic parameter remains positive which is
necessary for a realistic stellar configurations. We can conclude
our discussion as follows.
\begin{itemize}
\item The anisotropic stresses and energy density are positive throughout the star configurations.
\item The $r-$derivatives of density and anisotropic stresses
(i.e., density and pressure gradients) remains negative.
\item All types of ECs are valid.
\item The sound speeds remain within the bounds of $[0,1]$ (i.e, compact stars are stable).
\item The equation of state parameter lies between 0 and 1 for each star radius.
\item The measure of anisotropy remains positive at the star core.
\end{itemize}


\vspace{0.5cm}

\begin{thebibliography}{40}

\bibitem{zs1} S. Weinberg, Rev. Mod. Phys. \textbf{61}, 1 (1989);
P. J. E. Peebles and B. Ratra, Rev. Mod. Phys. \textbf{75}, 559 (2003);
V. Husain and B. Qureshi, Phys. Rev. Lett. \textbf{116}, 061302 (2016).

\bibitem{zs2} D. Pietrobon, A. Balbi, and D. Marinucci, Phys. Rev. D \textbf{74},
043524 (2006); T. Giannantonio et al., Phys. Rev. D \textbf{74}, 063520 (2006); A. G. Riess et al.,
Astrophys. J. \textbf{659}, 98 (2007).

\bibitem{zs3} A. Qadir, H. W. Lee, and K. Y. Kim, Int. J. Mod. Phys. D \textbf{26}, 1741001 (2017).

\bibitem{R2}
S. Capozziello and V.~Faraoni, \textit{Beyond Einstein Gravity}
(Springer, Dordrecht, 2010).

\bibitem{R3} S. Capozziello and M.~De Laurentis,
Phys.\ Rept.\ {\bf 509}, 167 (2011) [arXiv:1108.6266 [gr-qc]].
%
\bibitem{R4} K. Bamba, S.~Capozziello, S.~Nojiri and S. D.~Odintsov,
  Astrophys.\ Space Sci.\  {\bf 342}, 155 (2012)
  [arXiv:1205.3421 [gr-qc]];\

\bibitem{R5} K. Koyama,Rep. Prog. Phys. \textbf{79}, 046902 (2016)
  [arXiv:1504.04623 [astro-ph.CO]].\
%
\bibitem{R6} \'{A}. de la Cruz-Dombriz~and~D. S\'{a}ez-G\'{o}mez,
  Entropy {\bf 14}, 1717 (2012)
  [arXiv:1207.2663 [gr-qc]];\
%
\bibitem{R7} K. Bamba, S. Nojiri and S. D.~Odintsov,
  arXiv:1302.4831 [gr-qc];\

\bibitem{R8} K. Bamba and S. D. Odintsov,
   arXiv:1402.7114 [hep-th];\
%
  Symmetry {\bf 7}, 220 (2015)
  [arXiv:1503.00442 [hep-th]].
\bibitem{R9} Z. Yousaf, K.~Bamba~and~M. Z. Bhatti, Phys. Rev. D \textbf{93}, 064059 (2016) [arXiv1603.03175 [gr-qc]].

\bibitem{R10} Z. Yousaf, K.~Bamba~and~M. Z. Bhatti, Phys. Rev. D \textbf{93},
124048 (2016) [arXiv:1606.00147 [gr-qc]].


\bibitem{k3} S. Nojiri, and S. D. Odintsov, eConf C {\bf 0602061}, 06 (2006);
Int.\ J.\ Geom.\ Meth.\ Mod.\ Phys.\ {\bf 4}, 115 (2007)
[hep-th/0601213].
\bibitem{k4} S. Nojiri, and S. D. Odintsov, arXiv:0801.4843 [astro-ph] (2008).
\bibitem{k5} S. Nojiri, and S. D. Odintsov, arXiv:0807.0685 [hep-th] (2008).
\bibitem{k6} Sotiriou, T. P. and Faraoni, V.: Rev. Mod. Phys. \textbf{82}, 451 (2010).

\bibitem{no1} S. Nojiri, and S. D. Odintsov, Phys. Rev D \textbf{68}, 123512 (2003).

\bibitem{kkk1} Z. Yousaf, Eur. Phys. J. Plus \textbf{132}, 71 (2017).

\bibitem{z1fr} M. Sharif and~Z. Yousaf, Z.: Astrophys. Space Sci. \textbf{355}, 317 (2015);
M. Z. Bhatti and Z. Yousaf, Int. J. Mod. Phys. D \textbf{26},
1750029 (2017); ibid. Int. J. Mod. Phys. D \textbf{26}, 1750045
(2017); M. Z. Bhatti and Z. Yousaf, Eur. Phys. J. C \textbf{76}, 219
(2016) [arXiv1604.01395 [gr-qc]].

\bibitem{z2frt} T. Harko, F. S. N. Lobo, S. Nojiri and S. D. Odintsov,
Phys. Rev. D \textbf{84}, 024020 (2011); Z. Yousaf and M. Z. Bhatti, Eur. Phys. J. C \textbf{76}, 267
(2016) [arXiv:1604.06271 [physics.gen-ph]];
M. Sharif~and~Z. Yousaf,~Astrophys.~Space~Sci.~\textbf{354},~471~(2014); Yousaf, Z., ~Bamba,
K.~and~Bhatti,~M.~Z.:~Phys. Rev. D \textbf{95}, 024024 (2017)
[arXiv:1701.03067 [gr-qc]].

\bibitem{z3frtrmn} S. D. Odintsov and D. S\'{a}ez-G\'{o}mez, Phys. Lett. B, \textbf{725}, 437 (2013);
Z. Haghani, T. Harko, F. S. N. Lobo, H. R. Sepangi and S. Shahidi, Phys. Rev. D, \textbf{88}, 044023 (2013);
I. Ayuso, J. B. Jim\'{e}nez and \'{A}  de la Cruz-Dombriz, Phys. Rev. D, \textbf{91}, 104003 (2015);
Z. Yousaf, M. Z. Bhatti and U. Farwa, Class. Quantum Grav. \textbf{34}, 145002 (2017).

\bibitem{bin4} Nojiri, S., Odintsov, S. D. and  Oikonomou, V.K.: arXiv: arXiv:1705.11098 [gr-qc].

\bibitem{zs4} A. Sokolov, J. Exp. Theor. Phys. \textbf{79}, 1137 (1980).

\bibitem{zs5} R. Sawyer, Phys. Rev. Lett. \textbf{29}, 382 (1972)

\bibitem{zs6} L. Herrera, N. O. Santos, and A. Wang, Phys. Rev. D \textbf{78},
084026 (2008); L. Herrera, G. Le Denmat, and N. O. Santos, Phys. Rev. D
\textbf{79}, 087505 (2009).

\bibitem{zs6a} A. Di Prisco, L. Herrera, J. Ospino, N. O. Santos, and
V. M. Vi\~{n}a-Cervantes, Int. J. Mod. Phys. D \textbf{20}, 2351
(2011); Z.~Yousaf,~K.~Bamba~and~M.~Z.~Bhatti,~Phys. Rev. D
\textbf{95}, 024024 (2017) [arXiv:1701.03067 [gr-qc]];
Z. Yousaf and M. Z. Bhatti,  Eur. Phys. J. C \textbf{76}, 267 (2016) [arXiv:1604.06271 [physics.gen-ph]].

\bibitem{zs11} L. Herrera, A. Di Prisco and J. Ospino, Gen. Relativ. Gravit. \textbf{85}, 044022 (2012).

\bibitem{zs12} M. Sharif and Z. Yousaf, Astrophys. Space Sci. \textbf{357}, 49 (2015);
Z. Yousaf and M. Z. Bhatti, Mon. Not. R. Astron. Soc.
\textbf{458}, 1785 (2016) [arXiv:1612.02325 [physics.gen-ph]].

\bibitem{zs15} R. A. Sussman and L. G. Jaime, arXiv:1707.00191 [gr-qc].

\bibitem{zs17} H. Shabani and A. H. Ziaie, Eur. Phys. J. C \textbf{77}, 31 (2017).

\bibitem{zs18} R. Garattini and G. Mandanici, Eur. Phys. J. C \textbf{77}, 57 (2017).

\bibitem{zs19} P. K. Sahoo, P. Sahoo, B. K. Bishi,
Int. J. Geom. Meth. Mod. Phys, \textbf{14}, 1750097 (2017) [arXiv:1702.02469 [gr-qc]];
S. K. Sahu, S. K. Tripathy, P. K. Sahoo and A. Nath, Chin. J. Phys. \textbf{55}, 862 (2017).

\bibitem{zs20} S. W. Hawking and G. F. R. Ellis,
\emph{The Large Scale Structure of Spacetime} (Cambridge University Press, Cambridge, 1979).

\bibitem{zs23} L. Herrera, and N. O. Santos, Phys. Rep. \textbf{286}, 53 (1997); L. Herrera, A. Di Prisco,
J. R. Hernandez, and N. O. Santos, Phys. Lett. A \textbf{237}, 113 (1998);
M. Z. Bhatti, Eur. Phys. J. Plus \textbf{131}, 428 (2016);
M. Z. Bhatti, Z. Yousaf and S. Hanif, Eur. Phys. J. Plus \textbf{132}, 230 (2017).

\bibitem{zs24} G. J. Olmo, Phys. Rev. D \textbf{75}, 023511 (2007); F. Briscese and
E. Elizalde, Phys. Rev. D \textbf{77}, 044009 (2008); \'{A}. de la Cruz-
Dombriz, A. Dobado, and A. L. Maroto, Phys. Rev. D \textbf{80},
124011 (2009); T. Clifton, P. G. Ferreira, A. Padilla, and C. Skordis,
Phys. Rep. \textbf{513}, 1 (2012).

\bibitem{zs25} S. Capozziello, M. De Laurentis, S. D. Odintsov, and
A. Stabile, Phys. Rev. D \textbf{83}, 064004 (2011);
S. Capozziello, M. De Laurentis, I. De Martino, M. Formisano, and S. D. Odintsov, Phys. Rev. D \textbf{85}, 044022 (2012).

\bibitem{zs26} J. A. R. Cembranos, \'{A}. de la Cruz-Dombriz and B. M. N\'{u}\~{n}ez,
J. Cosmol. Astropart. Phys. \textbf{04}, 021 (2012).

\bibitem{zs31} M. Sharif and Z. Yousaf, Eur. Phys. J. C  \textbf{75}, 194 (2015) [arXiv:1504.04367v1 [gr-qc]];
\textbf{351}, 351 (2014); M. Sharif and Z. Yousaf, Int. J. Theor. Phys. \textbf{55}, 470 (2016);
M. Z. Bhatti and Z. Yousaf, Eur. Phys. J. C \textbf{76}, 219 (2016) [arXiv1604.01395 [gr-qc]];
Z. Yousaf, M. Z. Bhatti and U. Farwa, Mon. Not. R. Astron. Soc. \textbf{464}, 4509 (2017);
M. Z. Bhatti, Z. Yousaf and S. Hanif, Phys. Dark Universe \textbf{16}, 34 (2017).

\bibitem{zs40} J.-Q. Guo and P. S. Joshi, Phys. Rev. D \textbf{94}, 044063 (2016).

\bibitem{zs33} J. Santos, J. S. Alcaniz, M. J. Rebou\c{c}as and F. C. Carvalho, Phys. Rev. D \textbf{76},
083513 (2007).

\bibitem{zs34} J. Wang, Y.-B. Wu, Y.-X. Guo, W.-Q. Yang and L. Wang, Phys. Lett. B \textbf{689}, 133 (2010).

\bibitem{zs35} M. Shiravand, Z. Haghani and S. Shahidi, arXiv: 1507.07726v2 [gr-qc].

\bibitem{zs36} J. Santos, M. Reboucas and J. Alcaniz, Int. J. Mod. Phys. D \textbf{19} 1315 (2010)
[arXiv:0807.2443].

\bibitem{zs37} J. Wang and K. Liao, Class. Quantum Grav. \textbf{29}, 215016 (2012) [arXiv:1212.4656].

\bibitem{zs38} Z. Yousaf, M. Ilyas and M. Z. Bhatti, Eur. Phys. J. Plus \textbf{132}, 268 (2017).

\bibitem{zs39} K. Atazadeh and F. Darabi, Gen. Relativ. Gravit. \textbf{46}, 1664 (2014).

\bibitem{kk37} N. M. Garc\'{i}a, T. Harko, F. S. N. Lobo and J. P. Mimoso, Phys. Rev. D \textbf{83}, 104032 (2011);
S. Nojiri, S. D. Odintsov and P. V. Tretyakov, P. V.: Prog. Theor. Phys. Suppl. \textbf{172}, 81 (2008).

\bibitem{zs42} D. Shee, S. Ghosh, F. Rahaman, B. K. Guha and S. Ray, arXiv:1612.05109v2  [physics.gen-ph];
S. K. Maurya
and M. Govender, arXiv:1703.10037v1 [physics.gen-ph]; S. K. Maurya, Y. K. Gupta, S. Ray and
D. Deb, Eur. Phys. J. C \textbf{76}, 693 (2016) [arXiv:1607.05582 [physics.gen-ph]];
S. K. Maurya, Y. K. Gupta, S. Ray and B. Dayanandan, Eur. Phys. J. C  \textbf{75}, 225 (2015)
[arXiv:1504.00209 [gr-qc]].

\bibitem{zs43} K. D. Krori and J. Barua, J. Phys. A: Math. Gen. \textbf{8}, 508 (1975).

\bibitem{50s} S. Weissenborn, I. Sagert, G. Pagliara, M. Hempel and J. Schaffner-Bielich, Astrophys.
J. \textbf{740}, L14 (2011).

\bibitem{51s} F. Weber, Prog. Part. Nucl. Phys. \textbf{54}, 193 (2005).

\bibitem{52s} H. Bondi, Proc. R. Soc. London A \textbf{282}, 303 (1964); H. A. Buchdahl, Astrophys. J. 146,
275 (1966).

\bibitem{53s} M. K. Mak, P. N. D. Jr. and T. Harko, Mod. Phys. Lett. A \textbf{15}, 2153 (2000) [arXiv:gr-
qc/0104031].

\bibitem{zs44} A. A. Starobinsky, Phys. Lett. \textbf{91B}, 99 (1980).

\bibitem{zs45} G. Cognola, E. Elizalde, S. Nojiri, S. Odintsov, L. Sebastiani and S. Zerbini, Phys. Rev.
D \textbf{77}, 046009 (2008) [arXiv:0712.4017 [hep-th]].

\bibitem{zs46} K. Bamba, C.-Q. Geng and C.-C. Lee, J. Cosmol. Astropart. Phys. \textbf{08}
021 (2010) [arXiv:1005.4574 [astro-ph.CO]].

\bibitem{zs47} Z. Yousaf, Eur. Phys. J. Plus \textbf{132}, 276 (2017).

\bibitem{zs48} K. Bamba, M. Ilyas, M. Z. Bhatti and Z. Yousaf, Gen. Relativ. Gravit. \textbf{49}, 112 (2017) [arXiv: 1707.07386 [gr-qc].

\bibitem{zs49} L. Herrera, Phys. Lett. A \textbf{165}, 206 (1992).

\bibitem{ps1} M. Sharif and Z. Yousaf, J. Cosmol. Astropart. Phys. \textbf{06}, 019 (2014);
M. Sharif and Z. Yousaf, Astrophys. Space Sci. \textbf{354}, 431 (2014);
M. Sharif and Z. Yousaf, Gen. Relativ. Gravit. \textbf{47}, 48 (2015).

\end{thebibliography}
\end{document}